# SIMPLE Dark Matter Search Results

TA Girard[*], F. Giuliani, T. Morlat, M. Felizardo da Costa[c]

[a]*Centro de Física Nuclear, Universidade de Lisboa, 1649--003 Lisbon, Portugal*

J.I. Collar

*Department of Physics, University of Chicago, Chicago IL, 60637 USA*

D. Limagne, G. Waysand[b], J. Puibasset

*INSP (UMR 7588 CNRS), Universités Paris 7 & 6, 75251 Paris, France*

H.S. Miley

*Pacific Northwest National Laboratory, Richland, WA 99352  USA*

M. Auguste, D. Boyer, A. Cavaillou

[b]*Laboratoire Souterrain à Bas Bruit (Université de Nice-Sophia Antipolis),*
*84400 Rustrel--Pays d'Apt, France*

J.G. Marques[a], C. Oliveira, M. Felizardo, A.C. Fernandes, A.R. Ramos[a]

*Instituto Tecnológico e Nuclear, Estrada Nacional 10, 2686--953 Sacavém, Portugal*

R.C. Martins

[c]*Department of Electronics, Instituto Superior Técnico, Av. Rovisco Pais 1, 1049--001 Lisbon Portugal*

Abstract

We report an improved SIMPLE experiment comprising four superheated droplet detectors with a total exposure of 0.42 kgd. The result yields ~ factor 10 improvement in the previously-reported results, and -- despite the low exposure -- is seen to provide restrictions on the allowed phase space of spin-dependent coupling strengths almost equivalent to those from the significantly larger exposure NAIAD-CDMS/ZEPLIN searches.

---

[*]Corresponding author: email address: criodets@cii,fc.ul.pt (TA Girard)

The inability to discover baryonic matter sufficient to explain the observed dynamics of the universe has (for a number of decades) set the quest for weakly interacting massive particles (WIMPs). The search for this dark matter continues to be among the forefront efforts of experimental physics.

The coupling of WIMPs with matter may be either spin independent or dependent, depending on the composition of the WIMP itself. SIMPLE (Superheated Instrument for Massive ParticLe Experiments) is one of only two experiments to search for evidence of spin-dependent WIMPs using fluorine-loaded superheated droplet detectors (SDDs), the other being PICASSO [1]. The SDD is based on the nucleation of the gas phase by energy deposition in the superheated liquid, which must fulfill two conditions [2]: (i) the energy deposited must be greater than a thermodynamic minimum, and (ii) this energy must be deposited within a minimum thermodynamically-defined distance inside the droplet. The two conditions together require energy depositions of order ~ 150 keV/µm for SIMPLE, rendering the detector effectively insensitive to the majority of traditional detector backgrounds which plague more conventional dark matter search detectors.

In 2000, we reported [3,4] first exclusion limits from a prototype measurement involving a single 9.2 g active mass SDD module operated for 16 day. These results demonstrated the essential performance qualities of the detector, but were limited by statistics.

We here report new results from four modules of a seven module test which provide almost an order of magnitude improvement on the prototype result. Even at the low exposure level of 0.42 kgd, the result approaches those of other larger mass/exposure spin-dependent searches, and provides significant, complementary restrictions on the allowed phase space of spin-dependent coupling strengths. The results are equivalent to those recently reported by PICASSO [5] with a 2 kgd exposure, demonstrating the impact of high device radiopurity.

The detectors were fabricated in-house from *$C_2ClF_5$* (R-115) according to previously-described procedures [4]. These were installed in the GESA area of the LSBB laboratory [6]: the set was placed inside a thermally-regulated 700 liter water bath, surrounded by three layers of sound and thermal insulation, resting on a dual vibration absorber. To reduce ambient noise, a hydrophone was placed within the detector water bath, and a second acoustic monitor positioned outside the shielding. At 1500 mwe, the ambient neutron flux is primarily a fission spectrum from the rock, estimated at well-below $4 \times 10^{-5}$ n/cm$^2$s. The surrounding water bath additionally acts as a ~ 30 cm thick neutron moderator, further reducing any ambient neutron flux by at least two orders of magnitude.

A bubble nucleation is accompanied by an acoustic shock wave, which is detected by a piezoelectric transducer immersed in a glycerine layer at the top of the detector. The transducer signal was amplified a factor $10^5$; in the case of an event in any of the detectors, the temperature, pressure, and threshold voltage level for each device, plus its waveform trace and fast Fourier transform, were recorded in a Labview platform.

Table 1: Data results, without acoustic detection efficiency or background correction.

| Detector | Active Mass (g) | Efficiency | Filter Nº1 | | Filter Nº2 | |
|---|---|---|---|---|---|---|
| | | | 8.9ºC (evts/kgd) | 3.3ºC (evts/kgd) | 8.9ºC (evts/kgd) | 3.3ºC (evts/kgd) |
| 2 | 9.9 | 0.52 ± 0.10 | 278 ± 52.6 | 21.2 ± 12.2 | 179 ± 42.1 | 7.1 ± 7.1 |
| 4 | 10.8 | 1.11 ± 0.19 | 54.6 ± 22.3 | 25.9 ± 13.0 | 36.4 ± 18.2 | 25.9 ± 13.0 |
| 5 | 10.4 | 0.83 ± 0.17 | 66.2 ± 25.0 | 13.4 ± 9.5 | 28.4 ± 16.4 | 0 |
| 7 | 11.1 | 1.21 ± 0.21 | 558 ± 70.3 | 271 ± 41.3 | 407 ± 60.1 | 221 ± 37.3 |

In contrast to the temperature-ramping of the prototype measurement, the detectors were operated for 10.2 days at 8.9ºC (2.0 atm), and 14.3 days at 3.3ºC (1.9 atm). Additional measurements were performed at 14ºC in order to insure that all low rate devices were actually operating properly.

Because the detectors are manufactured above ground and transported 700 km to the LSBB under 4 atm of pressure at 0ºC, it is not uncommon that some devices suffer damage/degradation with respect to in-house fabrication and performance specifications. In the first analysis stage, the individual detectors were physically inspected for fractures (which lead to spontaneous inhomogeneous nucleations at the fracture sites), and their responses over the measurement period were monitored with respect to raw signal rate, threshold and differential temperature behavior, and pressure evolution. Three of the devices were rejected as a result of fractures and/or performances outside specification tolerances.

Since the WIMP interaction is weak, no two detectors should yield a WIMP signal coincident in time. The data record was anti-coincidence filtered on an event-by-event basis, with the criteria that (i) one and only one of the in-bath detectors had a signal, and (ii) no monitoring detector had a simultaneous signal. As seen in Table 1 (Filter Nº1), detectors 4 and 5 yielded rates roughly a third of the prototype rate; detector 7 in contrast yielded a rate higher than that of the prototype.

The frequency spectrum of the recorded events consists of pulses from the detector transducers, computer, power supply and LSBB electrical system, with the fast Fourier transform of the transducer signal comprising a well-defined frequency response with a primary harmonic at ~ 6 kHz. A second filtering was imposed in which only the previously filtered events with a primary harmonic between 5.5-6.5 kHz were accepted. The resulting rate reduction is also shown in Table 1 (Filter Nº2).

At 3.3ºC the reduced superheat $s = [(T - T_b)/(T_c - T_b)]$, where $T_c$, $T_b$ are the critical and boiling temperatures of the $C_2ClF_5$ respectively, is so low that WIMP sensitivity is negligible, and the results can be used to estimate a lower limit on the overall background rate. Following from the response studies of Ref. [4], this was conservatively assumed flat between the two temperatures, yielding an average difference of 31.1±14.6 evts/kgd in the fully filtered results.

At 1500 mwe, the ambient muon flux is ~ $10^{-2}$ muons/m$^2$s. The response of SDDs, of both small and large concentration, to X-rays, α-rays and cosmic-ray muons is well-studied [7,8], with the threshold for SDD sensitivity to these backgrounds occuring for a reduced superheat $s \geq 0.5$. SIMPLE devices operated at 8.9ºC ($s \sim 0.3$) are sufficiently below this threshold for these contributions to be neglected [9], and the predominant backgrounds are either α, neutron or α-induced recoils, or continuing, undiscriminated pressure microleaks in the detector capping [3,4] which lead to acoustic signals.

The α response of the SDDs was studied by diluting a 400 Bq liquid $^{241}$Am source into the matrix prior to gel setting. At 2 atm, the 5.5 MeV α and 91 keV recoiling $^{237}$Np daughter cannot induce nucleations at temperatures below 7.5 and -5ºC, respectively, leading to three regimes of background (the third, high temperature regime originating from high dE/dx Auger electron cascades following interactions of environmental gamma rays with Cl atoms in the refrigerant [3] begins as a sudden rise at 15ºC). Prior to extensive component purification, the spectrum in non-calibration runs had a close resemblance to that of the $^{241}$Am-diluted studies. Currently, the gelating agent, polymer additives and glycerol are purified using a pre-eluted ion-exchanging resin specifically suited for actinide removal. Each ingredient is pressure-forced through 0.2 μm filters to remove motes that might act as nucleation centers. The freon is single distilled; the water, double distilled. The presence of a radiocontamination, measured at $\leq 5 \times 10^{-5}$ pCi/g U via low-level spectroscopy, yields an overall background level of < 0.5 evts/kg freon/d. Radon contamination is low because of the 2 atm overpressure, water immersion, and short Rn diffusion lengths of the SDD construction materials (glass, metal). Air trapped in the detector during the *in situ* capping is shielded by a ~-2 cm thick glycerine layer on top of the active portion of the detector; the estimated Rn contribution, based on the measured site concentration [6], is less than 1 evt/kgd.

The response of smaller SDDs to various neutron fields has been studied extensively [8,9,10] and found to match theoretical expectations. The SIMPLE detector response to neutrons was investigated using monochromatic low energy neutron beams generated by filtering the thermal column of the Portuguese Research Reactor, and calibrated using a Am/Be source. The beam results, reported elsewhere [11], are in good agreement with thermodynamic calculations, and yield a minimum threshold recoil energy of 8 keV at 9ºC. The efficiency calibrations yield an average 73 ± 5% acoustic detection efficiency; the individual detector calibrations are shown in Table 1.

The metastability limit of a superheated liquid is described by homogeneous nucleation theory [12], which gives a limit of stability of the liquid phase at approximately 90% of the critical temperature for organic liquids at atmospheric pressure. Given an exponential decrease of the spontaneous nucleation rate with decreasing temperature by approximately three orders of magnitude per degree, at 9ºC this is entirely negligible.

As evident, all of the above contributions are significantly below the rates of Table 1. During the prototype phase, refrigerant-free 'dummy' modules yielded signals indistinguishable from bubble nucleation events [3]. These were found to arise from pressure microleaks through the plastic SDD caps of the submerged devices;

design modifications in the mechanical capping resulted in dummy device rates as much as a factor 10 less than those before modification. Nevertheless, even for detectors 4 and 5, the signal rate can be almost entirely attributed to undiscriminated microleaks.

Assuming the difference between the fully filtered measurements at 9ºC and 3ºC to be entirely WIMPS, the upper rate limit is 55 evts/kgd. Since this difference ($n$) is more probably a sum of background and microleak events, a 90% C.L. upper limit to the unobserved WIMP rate can be set by computing the expectation value of the total number of events, $\mu_{evts}$, such that the probability of observing at least $n$ events is 90%. Subtracting the expected number of background events (computed by maximum likelihood) from $\mu_{evts}$ yields an estimate of 24.0 evts/kgd (corrected for acoustic detection efficiency) for the expected, unobserved WIMP events. Note that simply assuming no WIMPs were detected would yield a factor 4.4 lower limit on the expected WIMP rate.

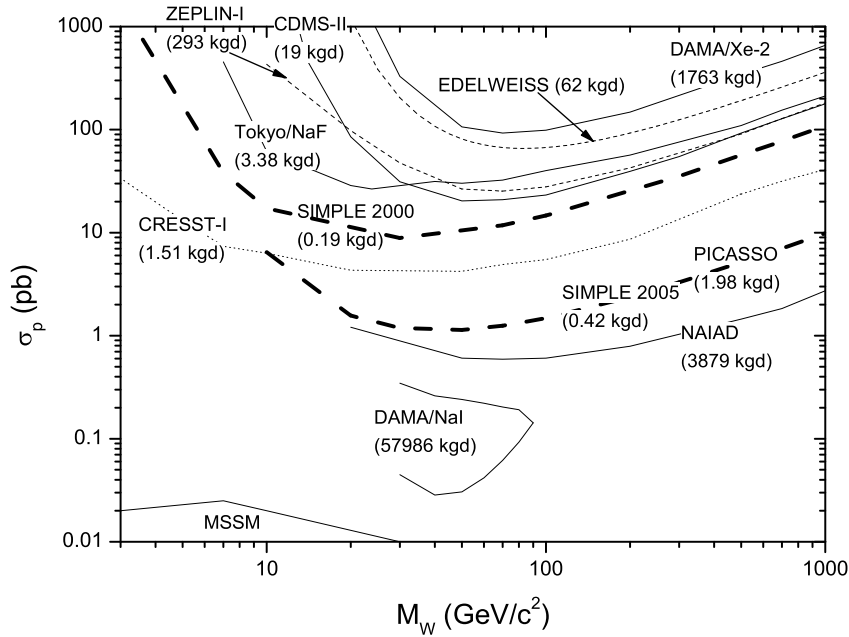

Fig. 1: Comparison of old and new SIMPLE limits, within the previous model-dependent translation to WIMP-proton exclusion plot with a standard halo model [13]: $v_{wimp}$ = 220 kms$^{-1}$, $v_{earth}$ in May = 257 kms$^{-1}$, and $\rho$ = 0.3 GeV/c$^2$cm$^3$. Odd N experiments are not shown since their main sensitivity is $\sigma_{Wn}$. The most recent PICASSO results are indistinguishable from those of SIMPLE 2005.

A comparison of the results with those of the SIMPLE prototype is shown in Fig. 1, using the cosmological parameters and method described in Ref. [13] in the calculation of the WIMP elastic scattering rates. The Figure indicates a level of 1.14 pb at 50 GeV/c$^2$, almost an order of magnitude improvement over the prototype result. A large part of this improvement results from the increased statistical level of measurement. The favorable comparison of SIMPLE with the larger exposure NAIAD [14] search is also evident from the Figure, clearly demonstrating the competitive power of the SDD technique in this application.

The constraints of Fig. 1 are obtained within the traditional model-dependent formulation based on the odd group approximation. The data were also analyzed using a model-independent formalism [15a,16], in which the spin-dependent interaction an be characterized in terms of either nucleons or coupling strengths; in the coupling strength representation, the cross section for a WIMP interaction with a nucleon is $\sigma_{SD} \sim [a_p<S_p> + a_n<S_n>]^2$, where $a_{p,n}$ ($S_{p,n}$) are the proton and neutron coupling strengths (proton and neutron group spins) respectively. Since the phase space is now 3 - dimensional ($a_p$, $a_n$, $M_W$), the results can be displayed by projection onto the $a_p$ - $a_n$ plane for a given $M_W$, as shown in Fig. 2 at 90% C.L. for $M_W$ = 50 GeV/c$^2$ (which is in the DAMA/NaI-preferred range [17]). Masses above or below this choice yield slightly increased limits. In both Figures, we use the spin values of Ref. [18]; use of the Ref. [19] values would lower the result of Fig. 1, rotating the SIMPLE and PICASSO curves in Fig. 2 about the origin to a more horizontal position.

Within this formulation, the region excluded by an experiment lies outside the indicated band, and the allowed region is defined by the intersection of the various bands. In this representation, the new SIMPLE result is already seen to eliminate a large part of the parameter space allowed by the significantly larger exposure Tokyo/NaF [20], NAIAD/NaI [14] and CRESST-I/Al$_2$O$_3$ [21] measurements at this mass cut, as well as the neutron-sensitive DAMA/Xe2 experiment [22]. The SIMPLE result is essentially equivalent to that of the most recent 2 kgd PICASSO report [5], but with an exposure of 0.42 kgd; the difference most likely results from the higher intrinsic backgrounds owing to the CsCl salts required in density-matching their gel and refrigerant.

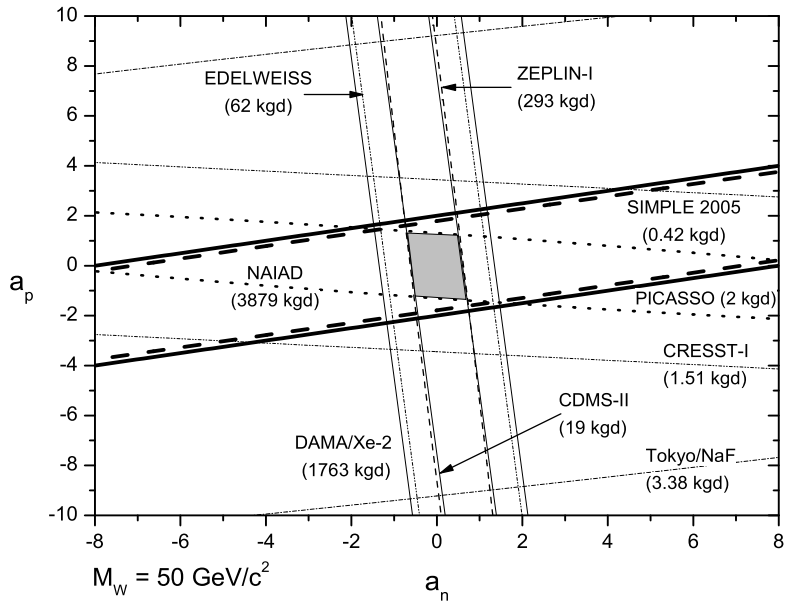

Fig. 2: $a_p$ - $a_n$ for SIMPLE (thick dashed), PICASSO (thick solid), NAIAD (dotted), CRESST-I and Tokyo/NaF for WIMP mass of 50 GeV/c$^2$. Also shown are the single nuclei DAMA/Xe2, and spin-INdependent EDELWEISS, CDMS and ZEPLIN-I. The region permitted by each experiment is the area inside the respective contour, with the shaded central region the allowed intersection of the NAIAD-CDMS/ZEPLIN measurements.

We also include the EDELWEISS [23], CDMS [24] and ZEPLIN-I [25], experiments, customarily considered as spin-INdependent searches, which by themselves are surprisingly even more efficient in reducing the allowed parameter space. The allowed area of the SIMPLE-CDMS/ZEPLIN intersection ($|a_p| \leq 2.4$, $|a_n| \leq 0.8$) at 50 GeV/c$^2$ is only slightly larger than that of NAIAD-CDMS/ZEPLIN ($|a_p| \leq 1.4$, $|a_n| \leq 0.7$). Equivalent limits in the model-independent cross section representation [15] are $\sigma_p \leq 0.7$ pb, $\sigma_n \leq 0.2$ pb.

The reasons for the large impact of the fluorine-based experiments are that (i) the relative sign of the fluorine $<S_n>/<S_p>$ is opposite to iodine, and (ii) both $<S_n>$ and $<S_p>$ of fluorine are non-negligible. Despite the small active detector mass, the limits reflect the favourable $^{19}$F spin structure, and the reduced background inherent to a detection method in which the sensitivity to several forms of background is effectively suppressed. Furthermore, the temperature-dependent threshold of the detector allows a background estimate from a measurement where the detector is no longer sensitive to neutralino-induced events. Note that identical limits obtain from only detectors 4 and 5 comprising a 0.21 kgd exposure, which themselves are still an order of magnitude above background estimates, with the difference most likely attributable to the continuing problem of microleaks.

In order to penetrate the frontier of the current allowed region of the $a_p$ - $a_n$ phase space shown in Fig. 2, only a modest 3 kgd exposure at the current SIMPLE performance is required. This would comprise seven devices of 10.5 g each operated over a period of ~ 40 days, which is relatively easily achievable. The recent application of pulse shape analysis techniques to the data records has moreover identified the possibility of discriminating between signal and microleak events, suggesting the ability to reach overall measurement rates of ~ 1 evt/kgd for the same 3 kgd exposure, corresponding to an ultimate factor of 10 further reduction in the exclusion. The SIMPLE project has recently received funding for conduct of the 3 kgd measurement.

At this level, further improvements will require implementation of clean room techniques towards increase in the device radiopurity. Given the current thrust of such searches to increasingly larger mass experiments, a hundredfold increase of the SIMPLE active mass to 10 kg (at a moderate cost of ~ US $100/kg) by modular construction would seem comparatively inexpensive and feasible.

**Acknowledgments**

We thank V. Zacek and C. Leroy for useful discussions, and J. Carneiro and N. Almeida for their contributions to the MCNP simulations of the filtered neutron beams. This work was supported in part by grants POCTI/FNU/43683/2002 and POCTI/FNU/32493/2000 of the Foundation for Science and Technology of Portugal, co-financed by FEDER.

**References**

[1]    L.A. Hamel et. al., Nucl. Instr. & Meth.  A388 (1997) 91.

[2]     F. Seitz, Phys. Fluids 1 (1958) 2.

[3]     J.I. Collar, J. Puibasset, TA Girard et. al., Phys. Rev. Lett. 85 (2000) 3083.

[4]     J.I. Collar, J. Puibasset, TA Girard et. al., New Journ. Phys. 2 (2000) 14.1.

[5]     M. Barnabé-Heider, M. Di Marco, P. Doane, et. al.: arXiv hep-ex/0502028 (2005); PICASSO is based on $C_3F_8$ and $C_4F_{10}$.

[6]     Laboratoire Souterrain à Bas Bruit de Rustrel-Pays d'Apt: http://lsbb.unice.fr; an electromagnetically shielded underground laboratory 60 km east of Avignon. The GESA (Génerateur Électrique des Servitudes Avancées) area at 1500 mwe constitutes a Faraday cage, reducing the magnetic field to less than 6 µT, with a long time stability of better than 20 nT and fluctuations below 2.5 fT√Hz. The radioactivity of the rock due to $^{136}$Cs is less than 0.437 Bq; to $^{226}$Ra, less than 0.645 Bq, with a radon average of 28 Bq/m$^3$.

[7]     N. Boukhira, I. Boussaroque, R. Gornea et. al., Astrop. Phys. 14 (2000) 227.

[8]     F. d'Errico, Nucl. Instr. & Meth. B142 (2001) 229.

[9]     M. Harper, PhD thesis, University of Maryland (1991).

[10]    Y.-Ch. Lo and R. Apfel, Phys. Rev. A38 (1988) 5260.

[11]    F. Giuliani, C. Oliveira, J.I. Collar, TA Girard et. al., Nucl. Instr. & Meth. A526 (2004) 348.

[12]    J. G. Eberhard et. al., J. Coll. Interf. Sci. 56 (1975) 369.

[13]    J. D. Lewin and P. F. Smith, Astrop. Phys. 6 (1996) 87.

[14]    B. Ahmed, G. J. Alner, H. Araujo et. al., Astrop. Phys. 19 (2003) 691.

[15a]   F. Giuliani, Phys. Rev. Lett. 93 (2004) 161301.

[15b]   F. Giuliani and T.A. Girard, Phys. Lett. B 588 (2004) 151.

[16]    D. R. Tovey, R.J. Gaitskell, P. Gondolo, Y. Ramachers and Roszkowski, Phys. Lett. B 488 (2000) 17.

[17]    R. Bernabei, M. Amato, P. Belli et. al., Phys. Lett. B 509 (2001) 197.

[18]    F. Pacheco and D. Strottman, Phys. Rev. D 40 (1989) 2131.

[19]    M. T. Divari, T. S. Kosmas, J. D. Vergados et. al., Phys. Rev. C61 (2000) 054612.

[20]    A. Takeda, M. Minowa, K. Miuchi et. al., Phys. Lett. B 572 (2003) 145.

[21]    W. Seidel, M. Altmann, G. Angloher et. al., 2002, in *Dark Matter in Astro- and Particle Physics Dark* (Springer-Verlag, Berlin, 2002).

[22]    R. Bernabei, P. Belli, F. Montecchia et. al., Phys. Lett. B 436 (1998) 379.

[23]    A. Benoit, L. Bergé, A. Broniatowski et. al., Phys. Lett. B 545 (2002) 43.

[24]    D.S. Akerib, J. Alvaro-Dean, M.S. Armel–Funkhouser et. al., Phys. Rev. Lett. 93 (2004) 211301.

[25]    V. A. Kudryavtsev and the Boulby Dark Matter Collaboration, in *Proc. of the 5th International Workshop on the Identification of Dark Matter* (2004).